\documentclass{jpsj-suppl}
\usepackage{txfonts} 
\usepackage{bm}

\newcommand {\pl}{\partial}

\newcommand {\al}{\alpha}
\newcommand {\be}{\beta}

\newcommand {\Ga}{\Gamma}

\newcommand {\la}{\lambda}
\newcommand {\La}{\Lambda}
\newcommand {\si}{\sigma}

\newcommand {\ep}{\epsilon}
\newcommand {\vep}{\varepsilon}
\newcommand {\e} {\mbox{e}}

\newcommand {\del}  {\delta}
\newcommand {\Del}  {\Delta}

\newcommand {\half}{ {\frac{1}{2}} }    

\newcommand {\Dcal}{{\cal D}}

\newcommand {\Itil}{{\tilde I}}

\newcommand {\rhotil} {{\tilde \rho}}

\newcommand {\Nbar}  {{\bar N}}

\newcommand {\Vbar}  {{\bar V}}

\newcommand {\rdot}{\dot{r}}

\newcommand {\xdot}{\dot{x}}

\newcommand {\bu}{{\bf u}}
\newcommand {\bv}{{\bf v}}
\newcommand {\bx}{{\bf x}}

\newcommand {\intx} {{\int dx}}

\newcommand {\ra} {\rightarrow}

\newcommand {\q}    {\quad}

\newcommand {\nl}    {\newline}

\newcommand {\PTP}  {Prog.Theor.Phys.}

\newcommand {\uinf} {{u^\infty}}
\newcommand {\finf} {{f^\infty}}
\newcommand {\tn} {{t_n}}
\newcommand {\un} {{u_n}}
\newcommand {\xn} {{x_n}}
\newcommand {\fn} {{f_n}}

\newcommand {\rhon} {{\rho_n}}
\newcommand {\alinv} {{\frac{1}{\alpha}}}
\newcommand {\sqral} {{\sqrt{\alpha}}}
\newcommand {\tauzero} {{\tau_0}}

\title{Velocity-Field Theory, Boltzmann's Transport Equation and Geometry  
}

\author{Shoichi \textsc{Ichinose}}

\inst{
Laboratory of Physics, School of Food and Nutritional Sciences, 
University of Shizuoka, 
Yada 52-1, Shizuoka 422-8526, Japan}

\email{ichinose@u-shizuoka-ken.ac.jp}

\recdate{July 15, 2013}

\abst{
Boltzmann equation describes the time development of 
the velocity distribution in the continuum fluid matter. 
We formulate the equation using the field theory where 
the {\it velocity-field} plays the central role. The matter (constituent particles) 
fields appear as the density and the viscosity.   
{\it Fluctuation} is examined, and is clearly discriminated from the quantum effect. 
The time variable is {\it emergently} introduced through the computational process step. 
The collision term, for  
the (velocity)**4 potential (4-body interaction), is explicitly obtained and 
the (statistical) fluctuation is closely explained. 
The present field theory model does {\it not} conserve energy and is an open-system model. 
(One dimensional) Navier-Stokes equation or Burger's equation,  appears. 
In the latter part, we present a way to directly define the 
distribution function by use of the geometry, appearing in the mechanical dynamics, 
and Feynman's path-integral. 
}

\kword{
Boltzmann equation, 
velocity field theory, 
statistical fluctuation, 
computational step number, 
open system, 
geometry 
}

\begin{document}
\maketitle

{\bf\large 1.\  Introduction}\q
  Boltzmann equation was introduced to explain the second law of 
the thermodynamics in the dynamical way, in 1872, by Boltzmann. 
We considers the (visco-elastic) fluid matter and examine the dynamical behavior using the velocity-field theory. 
The scale size we consider is far bigger than the atomic scale ($\sim 10^{-10}$m) and 
is smaller than or nearly equal to the optical microscope scale ($\sim 10^{-6}$m). 
The equation describes the 
temporal development of 
the distribution function $f(t,\bx,\bv)$ which shows the {\it probability} 
of a fluid-molecule (particle) having the velocity $\bv$ at the space $\bx$ and time $t$. 
\q We reformulate the Boltzmann equation using the field theory of the velocity 
field $\bu(\bx,~'t')$. Basically it is based on the {\it minimal energy principle}. 
We do {\it not} introduce time $t$. Instead of $t$, we use 
the {\it computational step number} $n$.  The system 
we consider consists of the huge number of fluid-particles (molecules) and the 
physical quantities, such as energy and entropy, are the {\it statistically-averaged} ones. 
It is not obtained by the deterministic way like the classical (Newton) mechanics. 
We introduce the {\it statistical ensemble} by 
using the well-established field-theory method, 
the {\it background-field method}. 
Renormalization phenomenon  occurs not from the {\it quantum} effect but from 
the {\it statistical fluctuation} due to 
the inevitable uncertainty caused by 1) step-wise (discrete-time) formulation and 
2) continuum formulation for the space. 

The dissipative system we consider is characterized by the dissipation of energy. 
Even for the particle classical (Newton) mechanics, the notion of energy is obscure 
when the dissipation occurs. We consider the movement of a particle 
under the influence of the friction force. The emergent heat (energy) during the period 
[t$_1$, t$_2$] can {\it not} be written as the following popularly-known form. \q
$\bm{<<}
\int_{x_1}^{x_2}F_{\mbox{friction}}~dx=\left[ E(x(t), \xdot(t)) \right]_{t_1}^{t_2}= 
E|_{t_2}-E|_{t_1}\ ,\ 
x_1=x(t_1)\ ,\ x_2=x(t_2)
\ ,\ (1)
\bm{>>}$
where $x(t)$ is the orbit (path) of the particle. It depends on the path (or orbit) itself. It cannot be written as the form of difference between some quantity at 
time t$_1$ and t$_2$. In this situation, we realize the time itself should be re-considered 
when the dissipation occurs.  
We have stuck to, due to Einstein's idea of "space-time democracy", 
the view that space and time should be treated on the equal footing. We 
present here the {\it step-wise} approach to the time-development. 

 We do {\it not} use time variable. Instead we use the computational-process step number $n$. 
Hence the {\it increasing} of the number $n$ is identified as the {\it time development}. 
The connection between step $n$ and step $n-1$ is determined by 
the {\it minimal energy principle}. In this sense, time is "emergent" from 
the minimal energy principle. The direction of flow (arrow of time) is built in from the beginning. 

{\bf\large 2.\q Emergent Time and Diffusion (Heat) Equation}\q
We consider 1 dimensional viscous fluid , and the velocity 
field $\{ u(x); -\infty <x<\infty \}$ describes the velocity distribution in the 1 dim space.  
Let us take the following energy functional[1] of the velocity-field u(x), 
$\bm{<<}
I_n[u(x)]=\intx \{ \frac{\si}{2\rhotil_0}(\frac{du}{dx})^2 +V(u)+u\frac{dV^1(x)}{dx}
+\frac{1}{2h}(u-u_{n-1})^2\}+I_n^0,\  
V(u)=\frac{m^2}{2}u^2+\frac{\la}{4!}u^4
,
\ (2)\bm{>>}$
where 
$u=u(x),\ u_{n-1}=u_{n-1}(x), \si\equiv 1,\  \rhotil_0\mbox{({\it mass-density})}\equiv 1,\  n=1,2,\cdots$. 
%
$I_n^0$ is a 'constant' term which is independent of $u(x)$. Later we will fix it. 
$\si$ is the {\it viscosity} constant and is also taken to be 1. 
$m^2$ is the mass density:\ (the mass of the fluid-particle)/ 2$\ell$. 
The quantity (2
) is the total energy of the fluid. 
The {\it velocity potential} $V(u)$ has the mass term and the 4-body interaction term. 
$V^1(x)$ is the (ordinary) position-dependent potential. 
${dV^1(x)}/{dx}$ is the external source (force) in this velocity-field theory. 
$h$ is some constant which will be 
identified as the {\it time-separation} for one step. 
$u_{n-1}(x)$ is taken to be a {\it given} velocity field at the (n-1)-th step.  
The n-th step velocity field  $u_n(x)$  
is given by the {\it minimal principle} of the n-th energy functional $I_n(u)$. 
This approach is callled "discrete Morse flows method". 

For simplicity we take the {\it periodic} boundary condition for the space:\ 
$
u(x)=u(x+2\ell)
$, 
where $2\ell$ is the periodic length. 
We may restrict the space region as $-\ell\leq x\leq \ell$. 
The variation equation $\del I_n(u)=0 (u(x)\ra u(x)+\del u(x))$ gives
$\bm{<<}\ 
\frac{1}{h}(u_n(x)-u_{n-1}(x))=\frac{\si}{\rhotil_0}\frac{d^2u_n}{dx^2}
-\frac{\del V(u_n)}{\del u_n}-\frac{dV^1_n(x)}{dx},\ 
\frac{\del V(u)}{\del u}=m^2u+{\la}/{3!}\cdot u^3
,\ 
(3)\bm{>>}$
where we have replaced the {\it minimal solution} by $u_n$. 
From the construction, we have the relation:\ 
$I_n[\un]\ \leq \ I_n[u_{n-1}]$. 
We, however, {\it cannot} say $I_n[\un]\ \leq\ I_{n-1}[u_{n-1}]$. 
The above equation 
describes the n-th step velocity field $u_n(x)$ in terms of $u_{n-1}(x)$ and 
vice versa. Hence it 
can be used for the {\it computer simulation}. 

We here introduce the {\it discrete time} variable $t_n$ 
as the step number n of $u_n$:\  
$
t_n=n h=n \tauzero\cdot h/\tauzero,\  \tauzero\equiv h\sqrt{\la\si}/m,\  n=0,1,2, \cdots
$, 
where $\tauzero$ is the time unit. 
The eq.(3
) is, in terms of the 'renewed' field $u(x, t)$,  expressed as
$\ \bm{<<}\ 
{1}/{h}\cdot (u(x, t_{n-1}+h)-u(x, t_{n-1}))=
{\pl^2u(x, t_n)}/{\pl x^2}-{\del V(u(x, t_n))}/{\del u(x, t_n)}
-{\pl V^1(x,t_n)}/{\pl x}
,
\ (4)\bm{>>}\ $
where we use $u(x, t_n)\equiv u_n(x),\ t_n=t_{n-1}+h$. 
As $h\ \ra\ 0$, we obtain
$
\pl u(x,t)/\pl t=\si/\rhotil_0\cdot\pl^2 u(x,t)/\pl x^2-\del V(u(x, t))/\del u(x, t)
-\pl V^1(x,t)/\pl x
$,\ 
where we have replaced both $t_n$ and $t_{n-1}$ by $t$. 
This is 1 dim {\it diffusion} equation with the potential $V(u)$.

We remind that the variational principle for the n-step energy functional $I_n[u(x)]$ 
\  (2
), 
$\del I_n=I_n[u+\del u]-I_n[u]=0$, 
gives 
$u_n(x)$ for the given $u_{n-1}(x)$. 
We regard the increase of the step number as the {\it time development}. 
Taking into account the fact that, at the (n-1)-step, the matter-particle 
at the point $x$ flows at the speed of $u_{n-1}(x)$, 
the energy functional $I_n$, (2
), should be replaced by the following one[1].  
$\ \bm{<<}\ 
\Itil_n[u(x)]=\intx \{ {\si}/{2\rhotil_0}\cdot ({du}/{dx})^2 +V(u)+u{dV^1(x)}/{dx}
+{1}/{2h}\cdot (u(x+hu_{n-1})-u_{n-1})^2\}+\Itil
_n^0.
\ (5)\bm{>>}\ $
Note that $u(x)-u_{n-1}(x)$ in eq.(2
) is replaced by $u(x+hu_{n-1}(x))-u_{n-1}(x)$. 
The step-wise recursion relation (3
) is corrected as
$\ \bm{<<}\ 
{1}/{h}\cdot (u_n(x)-u_{n-1}(x))+u_{n-1}(x) {d\un(x)}/{dx}
={\si}/{\rhotil_0}\cdot {d^2u_n}/{dx^2}
-{\del V(u_n)}/{\del u_n}-{dV^1_n(x)}/{dx}
.
\ (6)\bm{>>}\ $

  As done before, let us replace the step number $n$ by the {\it discrete} time $t_n=nh$. 
Taking the continuous time limit ($h\ \ra\ 0$), we obtain
$\ \bm{<<}\ 
\frac{\pl u(x,t)}{\pl t}+u(x,t)\frac{\pl u(x, t)}{\pl x}=\frac{\si}{\rhotil_0}\frac{\pl^2 u(x,t)}{\pl x^2}
-\frac{\del V(u(x, t))}{\del u(x, t)}
-\frac{\pl V^1(x,t)}{\pl x}\ .
\ (7)\bm{>>}\ $
This is called {\it Burgers's equation} (with the velocity potential $V(u)$ and the external 
force $\pl V^1/\pl x$) and is considered to be 1 dimensional 
Navier-Stokes equation. 
Note that the non-linear term in the LHS of eq.(7
) appears 
not from the potential ( velocity-field interaction) but from the {\it self-consistency}
of the velocity-field change from step $n-1$ to step $n$. 
${\pl}/{\pl t}+u {\pl}/{\pl x}\equiv{D}/{D t}$ is called 
{\it Lagrange derivative}. 

The equation (7
), for the massless case $m=0$, is invariant under 
the {\it global Weyl transformation}. 
\nl
$\ \bm{<<}\ 
V^1(x,t)\ra \e^{-2\vep}V^1(\e^\vep x, \e^{2\vep} t)\ ,\ 
u(x,t)\ra \e^{-\vep} u(\e^\vep x, \e^{2\vep}t)
\ ,\ 
\pl_x\ra\e^{-\vep}\pl_x\ ,\ 
\pl_t\ra\e^{-2\vep}\pl_t\ ,\ 
t\ra \e^{2\vep}t\ ,\ x\ra \e^\vep x
\ ,
\ (8)\bm{>>}\ $
where $\vep$ is the real constant parameter. 
 
For simplicity, we explain in one space-dimension (dim). The generalization to 2 dim 
and 3 dim is straightforward. 

{\bf\large 3.\q Statistical Fluctuation Effect}\q
We are considering the system of {\it large number of matter-particles}, 
hence the physical quantities, such as energy and entropy, are given by some 
{\it statistical average}. In the present approach, the system behavior $u_n(x)=u(x,t_n)$ is completely 
determined by eq.(6
) when the initial configuration $u_0(x)=u(x,0)$ is given. 
We have obtained the solution by the {\it continuous} variation $\del u(x)$ to $\Itil_n[u]$, (5
). In this sense, $u_n(x)$ is the 'classical path'. 
Here we should note that the present formalism is 
an {\it effective} way to calculate the physical properties of this {\it statistical} system. 
{\it Approximation} is made in the following points:\  
a1)\  
So far as $h\neq 0$, the {\it finite time-increment} gives {\it uncertainty} 
to  the minimal solution $u_n(x)$. 
This is because we cannot specify the minimum configuration definitely, 
but can only do it with {\it finite} uncertainty;\ 
a2)\ 
The real fluid matter is made of many micro particles with small but {\it non-zero size}. The existence 
of the characteristic particle size gives uncertainty to the minimal solution in 
this space-continuum formalism. 
Furthermore the particle size is not constant but does distribute in the statistical way. The shape of 
each particle differs. The present continuum formalism has limitation to describe the real situation 
accurately;\ 
a3)\ 
\ The system energy generally changes step by step. 
The present model (2
) describes an open-system. 
It means the present system {\it energetically} 
interacts with the outside. Such interaction is caused by the 
dissipative term in (2
). 

We claim the fluctuation comes {\it not} from the {\it quantum effect} but from the {\it statistics} due to 
the 
uncertainty which comes from 
the {\it finite} time-separation and the spacial distribution of {\it size} and {\it shape}. 
%
To take into account this fluctuation effect, we {\it newly} define the n-th energy functional $\Ga[u(x); u_{n-1}(x)]$ 
in terms of the original one $\Itil_n[u(x)]$, 
(5
), using the path-integral:\  
$\ \bm{<<}\ 
\e^{-\alinv\Ga[u(x); u_{n-1}(x)]}=\int\Dcal u(x)\e^{-\alinv \Itil_n[u(x)]}\nl
. (8b 
)
\bm{>>}\  
$
In the above path-integral expression, {\it all} paths $\{ u(x); -\ell\leq x\leq \ell \}$ are taken into account. 

We are considering the minimal path $u_n(x)$ as the dominant configuration 
and the small deviation $q(x)$ around it:\ 
$
u(x)=u_n(x)+ \sqral q(x),\ |\sqral q|\ll |u_n|,\ 
{\del \Itil_n[u]}/{\del u} |_{u=u_n}=0
$.\ 
Here a {\it new} expansion parameter $\al$ is introduced. 
([$\al$]=[$I_n$]=ML$^2$T$^{-2}$) 
As the above formula shows, 
$\al$ should be small. The concrete form should be chosen depending on problem 
by problem. It should {\it not} include Planck constant, $\hbar$, because the fluctuation does not 
come from the quantum effect. It should be chosen as:\ 
b1)\ the dimension is consistent;\ 
b2)\ it is proportional to the small scale parameter which characterizes 
the relevant physical phenomena such as the mean-free path of the fluid particle;\ 
b3)\ the precise value should be best-fitted with the experimental data. 

The background-field method tells us to do the Taylor-expansion around $\un$. \nl
$\ \bm{<<}\ 
\Itil_n[u(x)]=\Itil_n[u_n(x)+\sqral q(x)]=\sum_{l=0}^{\infty}\al^{l/2}\cdot {q(x)^l}/{l!}\cdot
({\del^l\Itil_n[u]}/{\del u(x)^l})|_{u_n}
=\sum_{l=0}^{\infty}S_l[u_n]
,\ {\del\Itil_n[u]}{\del u(x)}|_{u_n}\\
=0,
\e^{-\alinv \Ga[u_n(x); u_{n-1}(x)]}
=\e^{-\alinv \Itil_n[u_n(x)]}
\int\Dcal q\exp\left[\int dx
\{
-1/\al \cdot S_2
+O(q^3)
\}
                      \right],\ 
S_0=\Itil_n[u_n],S_1=\int dx \\ 
q(x) ({\del \Itil_n[u]}/{\del u})|_{u_n} =0
,
\alinv S_2
= \half\frac{d}{dx}(q\frac{dq}{dx})+\half q D q+O(h),\  
D\equiv -\frac{\si(=1)}{\rhotil_0(=1)}\frac{d^2}{dx^2}+\la {u_n}^2+m^2+\frac{1}{h}-\frac{du_{n-1}}{dx}
,
\ (9)\bm{>>}\ $
where we make the Gaussian(quadratic, 1-loop) approximation. 
\nl
$\ \bm{<<}\ 
\e^{-\alinv \Ga[u_n(x); u_{n-1}(x)]}=\e^{-\alinv \Itil_n[u_n(x)]}\times(\det D)^{-1/2},\ 
(\det D)^{-1/2}
=               \exp
\left\{
\half\Tr\int_0^\infty\frac{\e^{-\tau D}}{\tau}d\tau +\mbox{const}
\right\}
 ,
\ (10)\bm{>>}\ $
where $\tau$ is called Schwinger's proper time. ([$\tau$]=[$D^{-1}$]=L/M.)

We evaluate $\ln(\det D)^{-1/2}=\half\int_0^\infty d\tau \Tr G(x,y)/\tau=  
\half\int_0^\infty d\tau \int_{-\ell}^{\ell}dx~ G(x,x)/\tau$. Up to the first order of $\Vbar$, 
the result is given by
$\ \bm{<<}\ 
\sqrt{\ep\La/\pi}-{1}/{2\sqrt{\pi\ep\mu}}\cdot\int_{-\ell}^\ell dz \ep (\la\un(z)^2+m^2+{1}/{h}+{du_{n-1}(z)}/{dz}),
\ (11)\bm{>>}\ $                
where 
the {\it infrared cut-off} parameter $\mu\equiv \sqrt{\si}/\ell$ and  
the {\it ultraviolet cut-off} parameter $\La\equiv h^{-1}$ are introduced. 
$\ep^{-1}\equiv \si/\rhotil_0 =1$. 
We see the mass parameter $m^2$ shifts under the influence of the fluctuation. 
$\ \bm{<<}\ 
m^2\ \ra\ m^2+{\al}/{\sqrt{\pi\ep\mu}}\cdot\ep\la=m^2+\al\la\sqrt{{\ell\rhotil_0}/{\pi\si\sqrt{\si}}},
\ (12)\bm{>>}\ $

 The coupling $\la$ is also shifted by the O($\Vbar^2$) correction. 
The shift of these parameters 
corresponds to the {\it renormalization} in the field theory. 
In this effective approach, we have {\it physical} 
cut-offs $\mu$ and $\La$ which are expressed by the (finite) parameters appearing in the starting 
energy-functional. When the functional (5
)  (effectively) 
works well, all effects of the statistical fluctuation reduces to the simple shift of the original parameters. 
This corresponds to the renormalizability condition in the field theory.

{\bf\large 4.\q Boltzmann's Transport Equation}\q
We use, for simplicity, the original names for the shifted parameters. 
The step-wise development equation (6
),  with 
$\del V/\del u=m^2u+{\la}/{3!}\cdot {u}^3+u_{n-1}\cdot {d\un}/{dx}$, $V^1_n=0$, 
is written as
$\ \bm{<<}\ 
\frac{1}{h}(u_n(x)-u_{n-1}(x))=\frac{d^2u_n}{dx^2}
-m^2u_n-\frac{\la}{3!}{u_n}^3-u_{n-1}\frac{du_n}{dx}
\ \ \mbox{or}\ \ 
u_{n-1}(x)=\frac{u_n(x)-h\{ {d^2u_n}/{dx^2}
-m^2u_n-{\la}/{3!}\cdot {u_n}^3 \}}{1-h\cdot {du_n}/{dx}}
.
\ (13)\bm{>>}\ $
The latter form is convenient for the 'backward' recursive computation:\ $\un\ra u_{n-1}$. 
When the system reaches the {\it equilibrium state} after sufficient recursive computation ($n\gg 1$), 
we may assume $u_{n-1}(x)=u_n(x)\equiv \uinf (x)$. $\uinf (x)$ satisfies:\ 
$
{d^2\uinf}/{dx^2}-m^2\uinf-{\la}/{3!}\cdot{\uinf}^3-\uinf\cdot {d\uinf}/{dx}=0
$. 

We here introduce the {\it distribution function} $f_n(x, v)$ as 
the probability for the matter-point particle in the space interval $x\sim x+dx$ 
and the velocity interval $v\sim v+dv$, at the step $n$, is given by\ 
$
{1}/{\Nbar_n}\cdot f_n(x, v) dx dv
$, 
where $\Nbar_n$ is the total particle number of the system at the step $n$. 
Then the n-th {\it distribution} $\fn (x,v)$ and 
the {\it equilibrium distribution} $f^\infty(x,v)$ are introduced as 
$\ \bm{<<}\ 
\un(x)={1}/{\rho_n(x)}\cdot\int v \fn(x, v) dv,\ 
\un(x)\ra\uinf(x)\ \mbox{and}\ \fn(x,v)\ra\finf(x,v)\ \mbox{as}\ n\ra\infty
,
\ (14)\bm{>>}\ $
where $\uinf(x)$ is the {\it equilibrium} velocity distribution. 
$\rho_n(x)$ is the {\it particle number} density. 
The continuity equation is given by\ 
$
{1}/{h}\cdot (\rho_n(x)-\rho_{n-1}(x))+{d}/{dx}\cdot (\rhon(x)\un(x))=0
$. 

The recursion relation (13
) is expressed, in terms of the distribution functions, as\ 
$\ \bm{<<} 
{1}/{h}\cdot [ \\ 
f_n(x+hu_{n-1}(x), v)-f_{n-1}(x, v)]=
{\pl^2f_n(x,v)}/{\pl x^2}-m^2f_n(x,v)-{\la}/{3!}\cdot\fn (x, v) {\un(x)}^2,
\ (15)\bm{>>}\ $
where\ $
\un(x)={1}/{\rho_n(x)}\cdot\int v \fn(x, v) dv$. 
This is the {\it Boltzmann's transport equation} for the 2-body and 4-body 
velocity-interactions. 
We can express the step-wise expression (15
) in the continuous time $t$ 
form as in Sec.2. 
This is the {\it integrodifferential equation} for $f_n(x,v)$ 
when $\rho_n$ is a constant. 
The right hand side (RHS) is called {\it collision term}. 
We notice when we may replace $u_{n-1}$, in the LHS of eq.(15
), by $\un$, the above 
recursion relation determine the (n-1)th step distribution $f_{n-1}$ by the n-th step 
data, $\fn$ and $\un$. 

In the remaining sections, we present an alternative approach to the distribution function $f_n(x,v)$.

\begin{figure}[h]
\begin{minipage}{16pc}
\includegraphics[width=16pc]{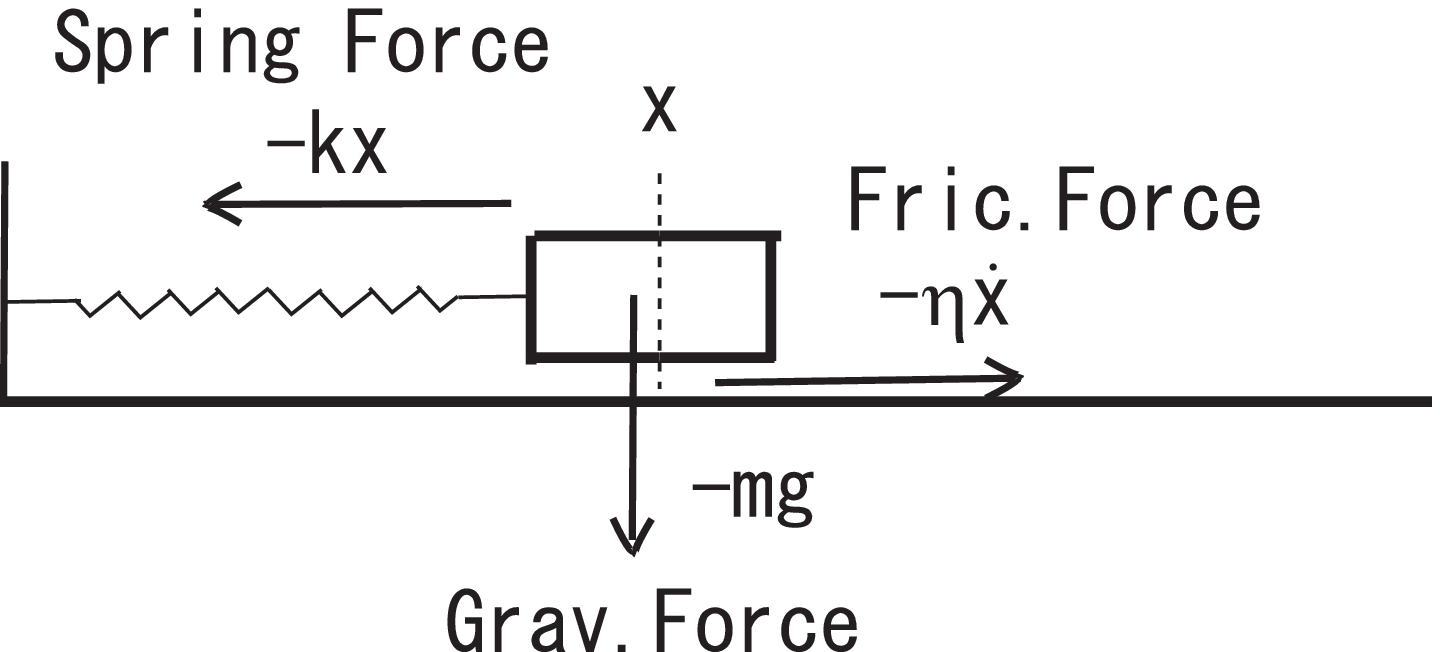}
\caption{\label{HOmodel}
The harmonic oscillator with friction, (17b). 
}
\end{minipage} 
\hspace{3pc}
\begin{minipage}{16pc}
\includegraphics[width=16pc]{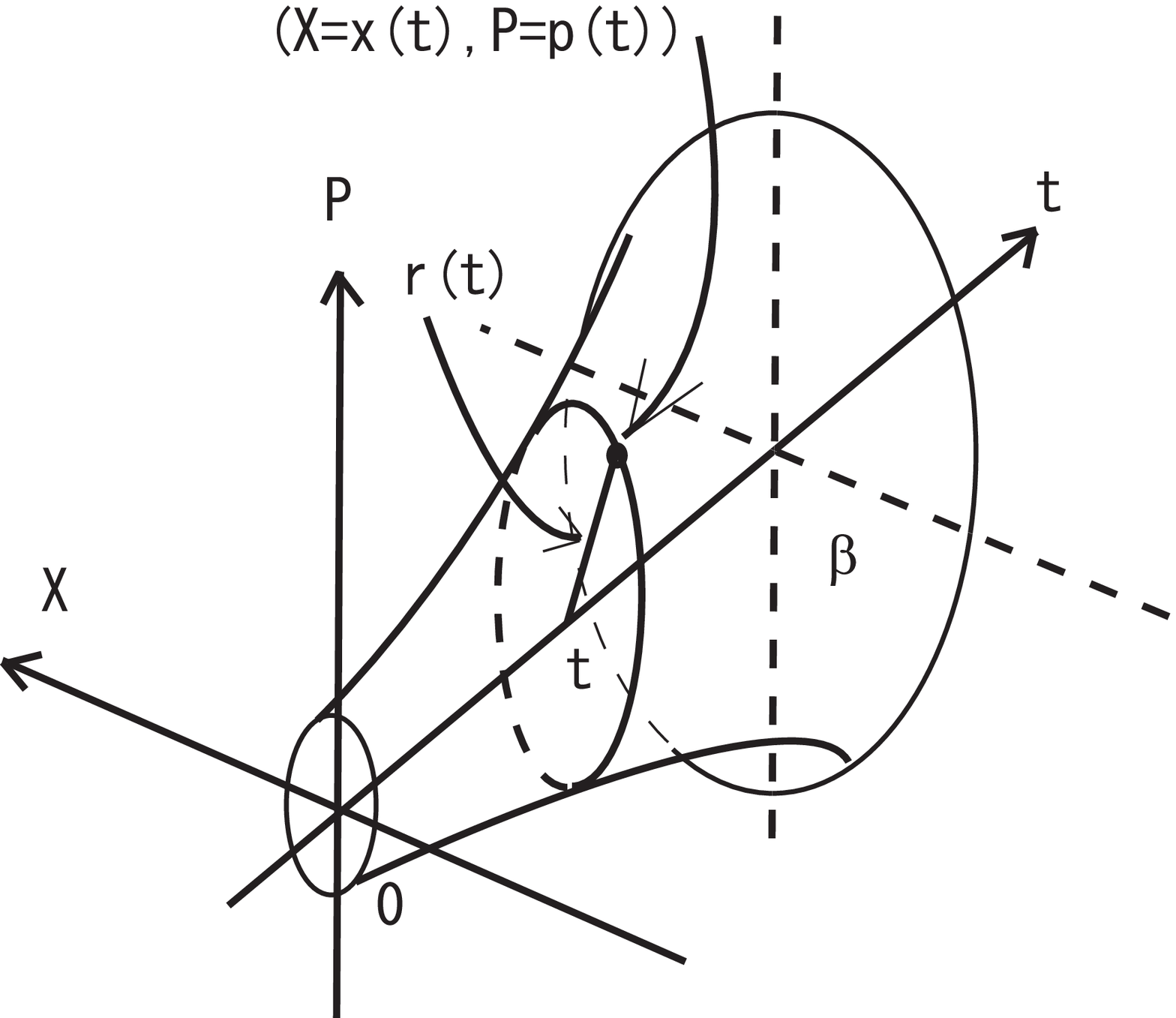}
\caption{\label{2DHyperSurf}
The two dimensional surface, (22b),  in 3D bulk space (X,P,t). 
}
\end{minipage}
\end{figure}

{\bf\large 5.\q Classical and Quantum Mechanics and Its Trajectory Geometry}\q
We can treat the classical mechanics and 
its quantization ( the quantum mechanics, not the quantum field theory) in the same way. 
In this case, 
the model is simpler than the previous case (space-field theory) and we can see 
the {\it geometrical structure} clearly. 
Let us begin with the energy function of a system variable , $x$, 
(1 degree of freedom). For example the {\it position} (in 1 dimensional space) 
of the harmonic oscillator with friction. 
We take the following $n$-th energy function to define the step flow. 
$\ \bm{<<}\ 
K_n(x)=  V(x)
+{\eta}/{2h}\cdot (x-x_{n-1})^2
+{m}/{2h^2}\cdot (x-2x_{n-1}+x_{n-2})^2+K_n^0
,
\ (16)\bm{>>}\ $
where $V(x)$ is the general potential and $K_n^0$ is a constant which does not depend on $x$. 
For the harmonic oscillator $V(x)=kx^2/2$ where $k$ is the spring constant. 
$\eta$ is the friction coefficient and $m$ is the particle mass. 
We assume $x_{n-1}$ and $x_{n-2}$ are given values. As in Sec.2, the n-th step position $\xn$ is 
given by the {\it minimal principle} of the n-th energy function $K_n(x)$:\ $\del K_n=0, x\ra x+\del x$. 
$\ \bm{<<}\ 
({\del V}/{\del x})|_{x=\xn}+{\eta}/{h}\cdot (\xn-x_{n-1})+{m}/{h^2}\cdot (\xn-2x_{n-1}+x_{n-2})=0
.
\ (17)\bm{>>}\ $
With the time $t_n$, the continuous limit ($h\ra 0$) gives us\q 
$\ \bm{<<}\ 
{dV(x)}/{dx}+\eta\cdot {dx}/{dt}+m\cdot {d^2x}/{dt^2}=0
\q (17b 
)\bm{>>}\ $, 
where $t_n=nh\ra t,\ \xn=x(t_n)\ra x(t),\ 
(\xn-x_{n-1})/h =dx/dt|_{t_n}\ra dx/dt,\ (\xn-2x_{n-1}+x_{n-2})/h =d^2x/dt^2|_{t_n}\ra d^2x/dt^2\ $. For the case of $V=kx^2/2$, this is the harmonic oscillator with the friction \ $\eta$. 
See Fig.\ref{HOmodel}. This is 
a simple {\it dissipative} system. 

  The recursion relation (17
) gives us, for the initial data $x_0$ and $x_1$, 
the series \{ $\xn =x(t_n) |n=0,1,2,\cdots$ \}. This is the classical 'path'. The fluctuation 
of the path comes from the {\it uncertainty principle} of the {\it quantum mechanics} in this case. 
(We are treating the system of 1 degree of freedom. No statistical procedure is necessary. ) 
As the time-interval $h$ tends to zero, the energy uncertainty grows ($\Del t\cdot \Del E\geq \hbar$). 
Hence the path $\xn$, obtained by the recursion relation (17
), has 
more uncertainty as $h$ goes to 0. 
As in Sec.3, we can generalize the n-th energy function $K_n(x)$, (16
), to 
the following one $\Ga(x_{n-1}, x_{n-2})$ in order to take into account the quantum effect. 
$\ \bm{<<}\ 
\e^{-\frac{1}{\hbar}h \Ga(x_{n-1}, x_{n-2})}=\int_{-\infty}^{\infty} dx~\e^{-\frac{1}{\hbar}h K_n(x)},
K_n(x)=  V(x)
+\frac{\eta}{2h}(x-x_{n-1})^2
+\frac{m}{2h^2}(x-2x_{n-1}+x_{n-2})^2
+K_n^0
.
\ (18)\bm{>>}\ $
We can evaluate the quantum effect by the expansion around 
the classical value $\xn$\ :\ $x=\xn +\sqrt{\hbar}~ q$ where $\hbar$ is Planck constant:\ 
$
\Ga_n\equiv\Ga(\xn; x_{n-1}, x_{n-2})=K_n(\xn)+{\hbar}/{2h}\cdot\ln(k+{\eta}/{h}+{m}/{h^2})
$, 
where the final expression is for the oscillator model: $V=kx^2/2$. The quantum effect 
does not depend on the step number $n$. It means the quantum effect contributes to the energy 
as an additional fixed constant at each step. 
The energy rate is obtained as
$\ \bm{<<}\ 
h\cdot {d\Ga(\tn)}/{d\tn}=\Ga_{n+1}-\Ga_{n}
=V(x_{n+1})-V(\xn)+
{\eta}/{2h}\cdot \{ (x_{n+1}-x_{n})^2 - (x_{n}-x_{n-1})^2 \}  
+{m}/{2h^2}\cdot \{  (x_{n+1}-2x_{n}+x_{n-1})^2  - (x_{n}-2x_{n-1}+x_{n-2})^2 \} 
K_{n+1}^0-K_n^0
.
\ (19)\bm{>>}\ $
The present system is again an open system, and the energy generally changes. 

In terms of the position difference $\xn-x_{n-1}\equiv \Del x_n$ and 
the velocity difference $(\xn-2x_{n-1}+x_{n-2})/h\equiv v_n-v_{n-1}\equiv \Del v_n$, 
we can rewrite the energy at $n$-step and 
read the {\it metric} as follows[2,3]. 
\nl
       $\ \bm{<<}\ 
K_n(\xn)=  V(\xn)
+{\eta}/{2h}\cdot (\xn-x_{n-1})^2
+{m}/{2h^2}\cdot (\xn-2x_{n-1}+x_{n-2})^2+K_n^0
={1}/{h^2}\cdot \{
V(\xn)(\Del t)^2
+{\eta h}/{2}\cdot (\Del \xn)^2
+{m h^2}/{2}\cdot (\Del v_n)^2
                   \}+K_n^0
,
        \ (20)\bm{>>}\ $
where $h$ (time increment) in the first term within the round brackets is replaced by $\Del t$. 
This shows us the metric for the n-step energy function is given by
\nl
$\ \bm{<<}\ 
(\Del s_n)^2\equiv 2h^2K_n(\xn)
=
2V(\xn'/\sqrt{\eta h})(\Del t)^2
+(\Del \xn')^2
+(\Del v_n')^2
 ,\ 
\xn'\equiv \sqrt{\eta h}\xn,\  {v_n}'\equiv \sqrt{mh^2}v_n
,
\ (21)\bm{>>}\ $
where , for the oscillator model, $V(\xn'/\sqrt{\eta h})=(k'/2){\xn'}^2, k'\equiv k/\eta h$. 
Eq.(21
) shows the energy-line element ${\Del s}^2$ in the ($t, \xn', {v_n}'$) space. 
Note that the above metric is {\it along the path}  
$x_n=x(t_n),\ v_n=v(t_n)=(x(t_n)-x(t_{n-1}))/h$ given by (17
). The metric is used, in the next section, as the {\it geometrical} basis for 
fixing the statistical ensemble.   

We take the freedom of the value $K_n^0$ in the following way. 
$\ \bm{<<}\ 
K_n^0=  -V(\xn)-{m}/{2h^2}\cdot (\xn-2x_{n-1}+x_{n-2})^2
+V(x_0)+{m}/{2h^2}\cdot (x_1-x_0)^2
.
\ (22)\bm{>>}\ $
This is chosen in such a way that the step $n$ energy $K_n(\xn)$, for the 
no dissipation ($\eta=0$), does {\it not} depend on the step number $n$ and 
the value is the total energy at the initial step (last 2 terms in (22
)). The graphs of 
movement and energy change can be plotted for various viscosities. 
For the no friction case,  the oscillator keeps the initial energy. 
When the viscous effect appears, the {\it energy changes step by step}, and finally 
reaches a constant {\it nonzero} value. 
We understand the finally-remaining energy (constant) as the dissipative 
one. Physically (in the real matter) the energy is realized as the pressure 
and the temperature which characterize the particle's "environment"(out-side world).  
For the resonate case ($4k/m\ =\ (\eta/m)^2$), both the movement and the energy 
are large.

{\bf\large 6.\q Statistical Ensemble, Geometry and Initial Condition}\q
In this section, we consider some statistical ensemble of the classical mechanical 
system taken in the previous section. Namely, we take $N$ 'copies' of the classical 
model and regard them as a set of  (1 dimensional) particles, where 
the dynamical configuration distributes in the probabilistic way. 
$N$ is a large number. 
The set has $N$ degrees of freedom:\ $x_1, x_2, \cdots, x_N$. 
As the physical systems, (1 dimensional) {\it viscous} gas and  {\it viscous} liquid are  examples. 
Each particle obeys the (step-wise) Newton's law (17
) with different 
{\it initial conditions}. 
$N$ is so large that we do not or can not observe the initial data. Usually we do not have interest 
in the trajectory of every particle and do not observe it. We have interest only in the macroscopic 
quantities such as {\it total energy} and {\it total entropy}. The N particles (fluid molecules) in the
present system are weakly interacting each other in such way that each particle almost independently  moves except that energy is exchanged. 
   
As the statistical ensemble, we adopt Feynman's idea of "path-integral"[2,3,4,5,6,7,8]
. 
We take into account all possible paths $\{ y_n \}$.  $\{ y_n \}$ need not satisfy (17
) nor 
certain initial condition. 
As the {\it measure} for the summation (integral) over all paths, 
we propose the following ones based on the geometry of (21
).  
Let us consider the following 2 dim surface in the 3 dim manifold ($X, P, t$):\ 
$\bm{<<}\ 
X^2+P^2=r^2(t),\ 0\leq t\leq \be
,\ (22b 
)\bm{>>}$\ 
where $r(t)$ is arbitrary (non-negative) function of $t$. We respect here 
the isotropy of the 2 dim phase space ($X, P$). See Fig.\ref{2DHyperSurf}. 
By varying the form of $\{ r(t):\ 0\leq t\leq \be\}$, we obtain different surfaces. 
Regarding each of them as a path used in Feynman's path-integral, we obtain 
the following statistical ensemble. First the {\it induced metric} $g_{ij}$ on the surface (22b) 
is given as
$\ \bm{<<}\ 
\left.(ds^2)_D\right|_{\mbox{on-path}}=\left. 2V(X)dt^2+dX^2+dP^2\right|_{\mbox{on-path}}
=\sum_{i,j=1}^{2}g_{ij}dX^idX^j,\ 
(g_{ij})=\left(
\begin{array}{cc}
1+{2V}/{r^2\rdot^2}\cdot X^2& {2V}/{r^2\rdot^2}\cdot X P \\
{2V}/{r^2\rdot^2}\cdot P X & 1+{2V}/{r^2\rdot^2}\cdot P^2
\end{array}
          \right)
,
\ (23)\bm{>>}\ $
where  $(X^1, X^2)=(X, P)$. Then the {\it area} of the surface (22b) is given by
$\ \bm{<<}\ 
A=\int\sqrt{\det g_{ij}}d^2X=\int\sqrt{1+{2V}/{\rdot^2}}dX dP
.
\ (24)\bm{>>}\ $
We consider all possible surfaces of (22b). The statistical distribution 
is, using the {\it area} $A$, given by 
$\ \bm{<<}\ 
\e^{-\be F}=\int_0^\infty d\rho\int {\begin{array}{c}
                                                        r(0)=\rho \\
                                                        r(\be)=\rho
                                                 \end{array}}
\prod_t\Dcal X(t)\Dcal P(t)\e^{-\frac{1}{\al}A}
,
\ (25)\bm{>>}\ $

In relation to Boltzmann's equation (Sec.4), we have directly defined 
the distribution function $f(t,x,v)$ using the geometrical quantities in the 
3 dim bulk space.

{\bf\large 7.\q Conclusion}\q
We have presented the field theory approach to Boltzmann's transport 
equation where the {\it velocity-field} distribution $u_n(x)$ plays the central role[9]. 
The collision term is {\it explicitly} obtained. 
Time is {\it not} used, instead the step number $n$ plays the role. 
We have presented the $n$-th energy functional (5
) which 
gives the step $n$ configuration $\un(x)$  
from the {\it minimal energy principle}. 
We regard the step flow ( the increase of $n$ ) as the evolution of 
the system, namely, {\it time-development}. 
Navier-Stokes equation is obtained by identifying time $t$ 
as $nh$.  
Time "emerges" and flows in a fixed direction. 
Fluctuation effect, due to the micro structure and 
micro (step-wise) movement, is taken into account by {\it generalizing} the $n$-th energy functional 
$\Itil_n[u(x)]$, (5
), to $\Ga[u(x); u_{n-1}(x)]$, (8b),  
where the classical path $\un(x)$ is dominant but {\it all possible paths} 
are taken into account (path-integral). {\it Renormalization} is explicitly done. The total 
energy generally does {\it not} conserve. The system is an open one, namely, the energy 
comes in from or go out to the outside. 
In the latter part we have presented a direct approach to the distribution function $f_n(x,v)$ 
based on the {\it geometry} emerging from the mechanical (particle-orbit) dynamics. 
We have examined the {\it dissipative} 
system using the {\it minimal (variational) principle} which is the key principle in the standard field theory. 
%
%
%
%
\nl
{\bf\Large References}\ [1]
N. Kikuchi, 
"Nematics-Mathematical and Physical Aspects", eds. J. -M. Coron, J. -M. Ghidaglia and 
H\'{e}lein, 
NATO Adv. Sci. Inst. Ser. C: Math. Phys. Sci. 332, Kluwer Acad. Pub., 1991, p195;\ 
[2]
S. Ichinose, J.Phys:Conf.Ser.\textbf{258}(2010)012003, arXiv:1010.5558;\  
[3] 
S. Ichinose, 
arXiv:1004.2573;\  
[4] 
S. Ichinose, 
\PTP\textbf{121}(2009)727, ArXiv:0801.3064v8;\ 
[5] 
S. Ichinose, 
ArXiv:0812.1263[hep-th];\  
[6] 
S. Ichinose, 
Int.Jour.Mod.Phys.24A(2009)3620
, arXiv:0903.4971;\ 
[7] 
S. Ichinose, 
J. Phys. :\ Conf.Ser.\textbf{222}
(2010)012048, 
ArXiv:1001.0222;\ 
[8] 
S. Ichinose, 
J. Phys. :\ Conf.Ser.\textbf{384}(2012)012028, 
ArXiv:1205.1316;\ 
[9] S. Ichinose, arXiv:1303.6616(hep-th)

\end{document}